\documentclass[nofootinbib,prdshowpacs,showkeys,preprintnumbers,twocolumn]{revtex4-1}
\usepackage{hyperref,amssymb,amsmath,mathrsfs,bm,graphicx}
\usepackage[dvipsnames]{xcolor}
\begin{document}

\title {Anisotropic star models in the context of vanishing complexity}
\author{C. Arias}
\affiliation{Perimeter Institute for Theoretical Physics \\
31 Caroline St. N. Waterloo Ontario, Canada, N2L-2Y5}

\author{E. Contreras }
\email{econtreras@usfq.edu.ec}
\affiliation{Departamento de F\'isica, Colegio de Ciencias e Ingenier\'ia, Universidad San Francisco de Quito,  Quito 170901, Ecuador\\}

\author{E. Fuenmayor}
\email{ernesto.fuenmayor@ciens.ucv.ve}
\affiliation{Centro de F\'isica Te\'orica y Computacional,\\ Escuela de F\'isica, Facultad de Ciencias, Universidad Central de Venezuela, Caracas 1050, Venezuela}

\author{A. Ramos}
\affiliation{School of Physical Sciences \& Nanotechnology,\\Yachay Tech University, 100119
Urcuqu\'{i}, Ecuador}

\begin{abstract}
We use the definition of complexity for static and self--gravitating objects to build up three physical general relativistic anisotropic models fulfilling the vanishing complexity condition which serves to provide the extra information needed to close the system of Einstein field equations. We evaluate the physical acceptability of these models by testing some of the conditions that the geometric and material sector must satisfy in order to be considered as reasonable realistic models. We present the results of this analysis by asserting that the studied cases demonstrate to be feasible and stable under the chosen set of parameters. Furthermore, the $P_{\perp}=0$ and the Consenza's anisotropy models that seem not satisfying the expect conditions are also discussed.
\end{abstract}


\maketitle

\section{Introduction}


General Relativity (GR) is, among all the theories proposed to describe the gravitational field, the one that owns the most acceptance in the scientific community. Einstein field equations (EFE) are the basis of this theory establishing a relationship between the space--time geometry and the energy content in space, represented by the Einstein's tensor and the energy--momentum tensor, respectively. To solve this system of differential equations inside the fluid configuration, we need to provide (in addition to the boundary or initial conditions) extra information related to the local physics. For example, it is well known that deviations of the isotropy and fluctuations of the local anisotropy in pressures may be caused by a large variety of physical phenomena of the kind  we expect to find in compact objects (see Refs. \cite{Herrera:1997plx}, for an extensive discussion on this point. See also, 
\cite{Ovalle:2017fgl,Ovalle:2017wqi,Ovalle:2019lbs,Gabbanelli:2019txr} for recent developments).   Besides, as  it  has  been  recently  proven, the isotropic pressure condition becomes unstable by the presence  of  dissipation,  energy  density  inhomogeneities and  shear \cite{Herrera:2020gdg}.  In this regard, in the case of a static spherically symmetric anisotropic fluid, EFE form a set of three ordinary differential equations for the five unknown functions (metric and matter hydrodynamics functions) so two extra conditions must be given \cite{herrera2008all,Abellan:2019jpq} in terms of an equation of state or an heuristic assumption involving metric and/or physical variables.  For instance, polytropic equations of state (see \cite{1chandrasekhar1939book, 2kippenhahn1990stellar} and references therein)
that were first considered in the Newtonian regime 
to study white dwarfs \cite{herrera2013newtonian,Abellan:2020ceu}, has been generalized to model
anisotropic spheres in General Relativity \cite{herrera2013general, 3abellan2020general, herrera2014conformally, tooper1964general, tooper1965adiabatic, tooper1966standard, bludman1973stability, nilsson2000general, maeda2002no, herrera2004evolution, lai2009polytropic, thirukkanesh2012exact}. Besides, physical conditions that restrict the metric variables have also been fully used such as the Karmarkar condition which allows to choose one of the metric functions as generator of the total solution \cite{karmarkar1948gravitational} (for recent developments see, \cite{Ramos:2021drk, singh2019generalized, tello2019anisotropic, ospino2020karmarkar, tello2020class, tello2020traversable, maurya2017anisotropic, ivanov2021generating, mustafa2020physically, ratanpal2020compact, pant2021new, shamir2020traversable, ashraf2020study, gomez2020charged, baskey2021analytical, maurya2021decoupling}, for example). Another example is the conformally flat condition which takes into account the vanishing of the Weyl tensor \cite{herrera2014conformally,Herrera:2001vg}. These studies show that different conditions and variables describe great variety of scenarios that characterize a determined behavior or evolution of the matter distribution for the self gravitating object. In this sense, a relevant and well known concept in physics, complexity, applied to the scope of General Relativity can perfectly be considered to provide the extra necessary information. As we shall see, this condition may be regarded as a non–local equation of state \cite{herrera2018new}, that can be used to obtain non-trivial configurations with zero complexity.\\

Complexity, rather than a simple and intuitive notion, is a physical concept that deeply relates to the fundamental structure of nature that has attracted a broad spectrum of researchers in various branches of science \cite{comp1, comp2, comp3, com4, comp5, comp6, comp7, comp8, comp9, comp10, comp11, comp12, comp13, grunwald2003kolmogorov, chapman2017complexity, alishahiha2015holographic, yang2020time}. Although so far there is not an unifying rigorous definition applied to all scenarios, several efforts have been made to provide a definition of complexity associated with the concepts of information and entropy \cite{comp11, comp12, comp13, grunwald2003kolmogorov, chapman2017complexity, alishahiha2015holographic, yang2020time}. Nevertheless, when considering self-gravitating objects a more natural definition of complexity is based on the structure of the fluid distribution (i.e. fluid inhomogeneity and pressure anisotropy), which is not directly related to information. In this regard, Herrera \cite{herrera2018new} proposed for the first time a complexity factor defined by means of a structure scalar (denoted by $Y_{TF}$) arising from the orthogonal splitting of the Riemann tensor \cite{bel1961inductions,herrera2009structure}, that manifestly exhibits that the complexity of a gravitational system is closely related to the internal structure of the object. This new approach of complexity \cite{herrera2018new}, defined for static spherically symmetric relativistic fluid distributions, stems from the basic assumption that one of the less complex systems corresponds to a homogeneous (in the energy density) fluid distribution with isotropic pressure, assigning a zero value of the complexity factor for such a configuration. Of course, as we shall see later, the vanishing complexity condition, $Y_{TF}=0$, can be achieved also for inhomogeneous and anisotropic self--gravitating spheres as demonstrated in  \cite{herrera2018new} and recently implemented in \cite{ospino2020karmarkar}.
In the present work, we intend to solve the system of EFE by using the vanishing complexity factor as a supplementary condition to construct three anisotropic models.\\

The work is presented as follows: In the next section we provide the equations, variables, conventions and conditions to deal with the system of Einstein equations. Section III is dedicated to presenting a brief revision of the definition of complexity for a static, spherically symmetric and anisotropic relativistic fluid. Next, a summary of the conditions for obtaining physically realistic compact star models will be displayed in section IV. In section V, three different models will be presented, analyzed and discussed in terms of its physical plausibility and relevance. Finally, a summary of the obtained results and some final remarks are presented in section VI.

\section{Einstein Field Equations}

In this section we present the relevant equations for describing a static, spherically symmetric  self--gravitating locally anisotropic fluids, which are bounded by a spherical surface $\Sigma$. In Schwarzschild--like  coordinates the metric reads as

\begin{equation}
ds^2=e^{\nu} dt^2 - e^{\lambda} dr^2 -
r^2 \left( d\theta^2 + \sin^2\theta d\phi^2 \right),
\label{metric}
\end{equation}
where $\nu$ and $\lambda$ are functions of $r$.  We number the coordinates: $x^0 =t$; $x^1 =r$; $x^2 =\theta$; $x^3 =\phi $. The metric (\ref{metric}) has to satisfy Einstein field equations
\begin{equation}
G^\nu_\mu=8\pi T^\nu_\mu,
\label{Efeq}
\end{equation}
where 
\begin{equation}
    T_{\mu \nu} = (\rho + P_{\perp})u_{\mu}u_{\nu} - P_{\perp}g_{\mu \nu} + (P_r - P_\perp) s_\mu s_{v}, 
\end{equation}
with $u^\mu$ the four velocity given by,
\begin{equation}
    u^\mu = (e^{-\nu/2},0,0,0),
\end{equation}
and $s^\mu$ is defined as,
\begin{equation}
    s^{\mu} = (0,e^{-\lambda/2},0,0),
\end{equation}
satisfying $s^\mu u_\mu = 0,$ $s^\mu s_\mu=-1$. Note that, 
\begin{equation}
    T_0^0 = \rho,\quad T_{1}^1 = -P_r,\quad T_2^2=T_3^3=-P_\perp \label{EMTensor},
\end{equation}
and this interpretation is unambiguous in the static case \cite{Bondi}. Therefore, the Einstein field equations read:
\begin{equation}
\rho =-\frac{1}{8\pi}\left[-\frac{1}{r^2}+e^{-\lambda}
\left(\frac{1}{r^2}-\frac{\lambda'}{r} \right)\right],
\label{fieq00}
\end{equation}
\begin{equation}
P_r =-\frac{1}{8\pi}\left[\frac{1}{r^2} - e^{-\lambda}
\left(\frac{1}{r^2}+\frac{\nu'}{r}\right)\right],
\label{fieq11}
\end{equation}
\begin{eqnarray}
P_\bot = \frac{1}{32\pi}e^{-\lambda}
\left(2\nu''+\nu'^2 -
\lambda'\nu' + 2\frac{\nu' - \lambda'}{r}\right),
\label{fieq2233}
\end{eqnarray}
where  primes stand for derivatives with respect
to $r$.\\
From the $r$ component of the conservation law
\begin{equation}
    \nabla_\mu T^{\mu \nu} = 0,
\end{equation}
one obtains the generalized Tolman-Opphenheimer--Volkoff (hydrostatic equilibrium) equation for anisotropic matter which reads
\begin{equation}
P'_r=-\frac{\nu'}{2}\left(\rho + P_r\right)+\frac{2\left(P_\bot-P_r\right)}{r}.\label{Prp}
\end{equation}
Alternatively, using 
\begin{equation}
\nu' = 2 \frac{m + 4 \pi P_r r^3}{r \left(r - 2m\right)},
\label{nuprii}
\end{equation}
we may write (\ref{Prp}) as,
\begin{equation}
P'_r=-\frac{(m + 4 \pi P_r r^3)}{r \left(r - 2m\right)}\left(\mu+P_r\right)-\frac{2\Pi }{r},\label{ntov}
\end{equation}
where $m$ is the  mass function defined by
\begin{equation}
R^3_{232}=1-e^{-\lambda}=\frac{2m}{r},
\label{rieman}
\end{equation}
or, equivalently
\begin{equation}
m= 4\pi \int_{0}^{r} \tilde{r}^{2} \rho d\tilde{r},
\label{m2}
\end{equation}
and, $\Pi$ is the local pressure anisotropy
\begin{equation}
\Pi = P_r - P_\perp.
\label{Pi}
\end{equation}

\noindent The exterior description of the spacetime is given by the Schwarzschild exterior solution \cite{schwarzschild1916gravitationsfeld}, 

\begin{equation}
ds^2= \left(1-\frac{2M}{r}\right) dt^2 - \frac{dr^2}{ \left(1-\frac{2M}{r}\right)} -
r^2 \left(d\theta^2 + \sin^2\theta d\phi^2 \right).
\label{SE}
\end{equation}

\noindent
Moreover, the matching conditions require continuity of the first and second fundamental form across the boundary $r=r_{\Sigma} = constant$, implying
\begin{equation}
e^{\nu_\Sigma}=1-\frac{2M}{r_\Sigma},
\label{enusigma}
\end{equation}
\begin{equation}
e^{-\lambda_\Sigma}=1-\frac{2M}{r_\Sigma},
\label{elambdasigma}
\end{equation}
\begin{equation}
[P_{r}]_{\Sigma}=0,
\label{PQ}
\end{equation}
where the subscript $\Sigma$ indicates that the quantity is
evaluated on the boundary surface $\Sigma$. Eqs. (\ref{enusigma}), (\ref{elambdasigma}), and (\ref{PQ}) are the necessary and sufficient conditions for a smooth matching of the two metrics (\ref{metric}) and (\ref{SE}) on $\Sigma$.

\section{Complexity Factor}
In the study of the complexity of static self-gravitating objects, it has been established that the simplest system (therefore having zero complexity) corresponds to a homogeneous and isotropic fluid. Hence, one expects that a complexity factor should measure the relation between the inhomogeneity in the energy density and the pressure anisotropy of a system. In this regard, Herrera \cite{herrera2018new} proposed one of the structure scalars obtained by the orthogonal splitting of the Riemann Tensor \cite{bel1961inductions, herrera2009structure}, as a candidate for the complexity factor for static self--gravitating fluids with spherical symmetry. Although the details of how these scalars were obtained is out of the scope of this paper, in this section we will briefly summarize the procedure related to obtaining the complexity factor $Y_{TF}$ proposed in \cite{herrera2018new} in order to achieve a self-contained work. For more information related to the structure scalars and the orthogonal splitting of the Riemann tensor, see references \cite{herrera2018new, bel1961inductions, herrera2009structure}. Let us define the tensor $Y_{\alpha\beta}$ by
 \begin{equation}
    Y_{\alpha\beta} = R_{\alpha\mu\beta\nu}\,u^{\mu}u^{\nu}
    \label{Y1},
 \end{equation}
which may be expressed as,
 \begin{equation}
    Y_{\alpha\beta} = \frac{1}{3} Y_T h_{\alpha\beta} + Y_{TF} \left(s_{\alpha}s_{\beta} + \frac{1}{3}h_{\alpha\beta}\right),
    \label{Y2}
 \end{equation}
with $R^{\mu}_{\alpha\beta\gamma}$ the components of the Riemman tensor and
\begin{eqnarray}
  h_{\mu\nu}&=&g_{\mu\nu}- u_{\mu}u_{\nu},
    \label{chi}\\
    Y_{T} &=& 4\pi(\rho + 3 P_r - 2 \Pi), \label{yt}\\
    Y_{TF} &=&4\pi \Pi + E = 8\pi \Pi - \frac{4\pi}{r^3} \int_{0}^r \tilde{r}^3 \rho' d\tilde{r}.\label{ytf}
 \end{eqnarray}
$\Pi$ is given by (\ref{Pi}) and $E$ is related to the electric part of the Weyl Tensor, defined as
 \begin{equation}
    E = - \frac{e^{-\lambda}}{4}\left[\nu'' + \frac{\nu'^2 - \lambda'\nu'}{2}- \frac{\nu' -\lambda'}{r} + \frac{2 (1 - e^{-\lambda})}{r^2} \right].
    \label{weyl}
 \end{equation}

 Note that these scalars are related to physical quantities such as energy density, energy density inhomogeneity (density contrast), and local anisotropy of pressure. In particular, the $Y_{TF}$ scalar (see Eq. (\ref{ytf})) relates the energy density inhomogeneity with local anisotropies in the pressure of the gravitating object. Furthermore, it is clear that $Y_{TF}$ vanishes for the simplest self--gravitating object (i.e. homogeneous and isotropic) as expected. Further, let us refer to the active gravitational (Tolman) mass \cite{tolman1930use}, which describes the energy content of a fluid \cite{tolman1930use}. The general Tolman mass for spherically symmetric static distributions of matter is given by,

\begin{equation}
m_T = 4\pi \int_{0}^{r_{\Sigma}} \tilde{r}^2 e^{(\nu+\lambda)/2} (T_{0}^0 - T_1^1 - 2 T^2_2) d\tilde{r}, \label{tolman1}
\end{equation}
which after replacing (\ref{EMTensor}) and some manipulation it can be obtained the following alternative expression,
 \begin{eqnarray}
 m_T &=& (m_T)_{\Sigma}\left(\frac{r}{r_{\Sigma}}\right)^3\nonumber\\
 &+& r^3 \int_{r}^{r_\Sigma} e^{(\nu+\lambda)/2}
 \left[ \frac{8\pi}{\bar{r}} \Pi - \frac{1}{\bar{r}^4} \int_{0}^{\bar{r}} 4 \pi \tilde{r}^3 \rho' d\tilde{r} \right]d\bar{r}.\nonumber\\ \label{tolmanM}
 \end{eqnarray}
Comparing (\ref{tolmanM}) with (\ref{ytf}) we may write
\begin{equation}
   m_T = (m_T)_{\Sigma}\left(\frac{r}{r_{\Sigma}}\right)^3 + r^3\int_{r}^{r_\Sigma}\frac{e^{(\nu+\lambda)/2}}{\bar{r}} Y_{TF} d\bar{r}\label{Mytf}.
\end{equation}
Eq. (\ref{Mytf}) shows that the scalar function $Y_{TF}$ encodes the influence of the local anisotropy of pressure and density inhomogeneity on the active gravitational  mass with respect to its value for the homogeneous isotropic fluid. In summary, $Y_{TF}$ is a quantity that relates the inhomogeneities in the energy density and the pressure anisotropies of the fluid, which vanishes for the simplest static self--gravitational system. \\

\noindent It is worth mentioning that from (\ref{ytf}) the homogeneous and isotropic system is not the only one with vanishing complexity parameter. Indeed, those systems are characterized to satisfy,

\begin{equation}
    \Pi = \frac{1}{2 r^3}\int_{0}^r \tilde{r}^3 \rho' d \tilde{r}, \label{vanishingYTF}
\end{equation}
which implies the existence of a broad family of solutions that could satisfy this condition. Furthermore, it provides a non local equation of state to solve the Einstein field equations (\ref{Efeq}), so it will be used as a condition to close the system in the construction of the models presented in this work.\\ 

\section{CONDITIONS FOR PHYSICAL REALISTIC MODELS}
\noindent The strategies to find solutions of Einstein field equations that serve to model real compact objects have been numerous. However, these solutions are no longer relevant if they are not physically acceptable. Therefore, modeled objects must meet acceptability conditions to be of physical interest \cite{delgaty1998physical}. These conditions have been identified over the years and compiled in \cite{Ivanov_2017}. We study the physical acceptability of anisotropic compact star models with vanishing complexity factor based on the most relevant conditions exposed in \cite{Ivanov_2017}. Here we give a briefly description and highlight them:

\begin{enumerate}
    \item The metric potentials must be positive, finite and free from singularities in the star's interior. Moreover, at the centre, they should satisfy $e^{-\lambda(0)}=1$ and $e^{\nu(0)}=const$. This fact, together with the joint conditions on the surface object, ensures that the compactness, $\frac{M}{R}$, must always be less than one half ($1/2$) in order to avoid singularities within the star.
    \item The density and the pressures must be positive inside the star and finite at the centre: $\rho(0)=\rho_{c}$, $P_{r}(0)=P_{rc}$, $P_{\perp}(0)=P_{\perp c} $. For $\rho$ this coincides with the null energy condition (NEC).
    \item The matter  sector should reach a maximum at the centre, so $\rho'(0)=P'_{r}(0)= P'_{\perp}(0)=0$ and should decrease monotonously outwards, $\rho'\leq 0$, $P'_{r} \leq 0$, $P'_{\perp}\leq 0$. This is consistent with stellar models that are in hydrostatic equilibrium.     
    \item The constraints on the energy-moment tensor, which must hold to describe a reasonable material configuration, are known as energy conditions. Regarding the energy conditions, a solution must satisfy the Dominant Energy Condition (DEC), which implies that $\rho$ $\geq$ $p_{r}$ and $\rho$ $\geq$ $P_{\perp}$. This condition imposes that the speed of the energy flow is always less than the speed of light (causality).
    \item The interior redshift, $Z=e^{-\nu/2}-1$, should decrease with the increase of $r$, since this depends only on the metric function $\nu$. On the surface, redshift and compactness are related in a simple way.
    \item Stability against cracking of perturbed self-gravitation spheres. This concept was introduced by Herrera et al. \cite{Herrera:1992lwz, di1994tidal, di1997cracking, Abreu_2007} to describe the behavior of self-gravitating anisotropic configurations just after departure from  hydrostatic equilibrium, when the total radial force changes its sign at some point within the fluid setting. According, to avoid cracking the configuration must satisfy:
    \begin{equation}
        -1 \leq \frac{dP_{\perp}}{d\rho}-\frac{dP_{r}}{d\rho} \leq 0.
    \end{equation}
\end{enumerate}

\section{Models}
In reference \cite{herrera2018new}, Herrera suggested some possible physical models to be analyzed and among all of them we may mention three which shall be developed here, namely:\\
\begin{enumerate}
\item[a] Like--Florides model, $P_{r}=0$
\item[b] A polytropic equation of state (proposed and partially developed in \cite{herrera2018new})
\item[c] A non-local equation of state in \cite{Hernandez_2004}
\end{enumerate}
In what follows, we shall explore each model in detail.
\subsection{Model I: Vanishing-Complexity $P_{r}=0$ case}
The first model into consideration describes a self-gravitating compact object with null radial pressure. This recalls the spherically symmetric solution that is held only by tangential pressure due to Florides \cite{florides1974new}, but unlike this case our model would not be homogeneous in density. Note that from (\ref{fieq11}) and $P_r=0$, we can integrate analytically and obtain a solution for the metric variable $\lambda$,
\begin{equation}
    e^{\lambda} = 8 \pi  r^2 \left(\frac{1}{8 \pi  r^2}+\frac{\nu '}{8 \pi  r}\right).\label{solM1}
\end{equation}

Then, after replacing (\ref{fieq00}), (\ref{fieq2233}), (\ref{Pi}) and (\ref{solM1}) into (\ref{vanishingYTF}), we obtain the following second order differential equation for $\nu$,
\begin{equation}
\left(r \nu '+2\right) \nu ''+\nu ' \left(r \nu ' \left(r \nu '-2\right)-2\right)=0,
\end{equation}
which must be integrated numerically. 

\begin{figure}[ht!]
\centering
\includegraphics[scale=0.5]{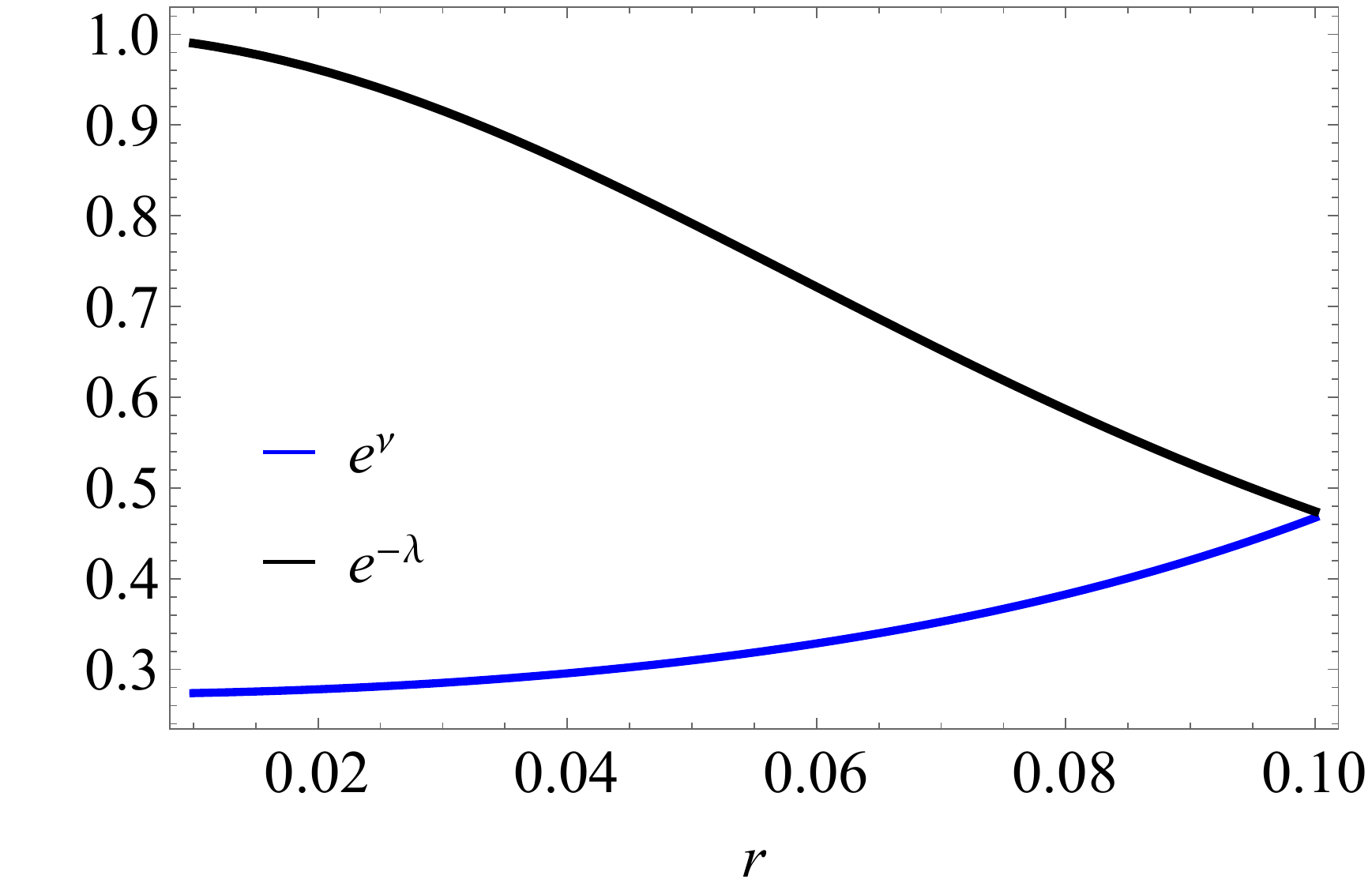}
\caption{\label{metricasprcero}
Metric functions $e^{\nu}$ (blue line) and $e^{-\lambda}$ (black line) as a function of the radial coordinate $r$ for the vanishing-complexity and $P_r =0$ model.
}
\end{figure}

In figure \ref{metricasprcero} we show the metric functions $e^\nu$ and $e^{-\lambda}$ as a function of the radial coordinate $x^{1}=r$. Note that both are positive, finite and free of singularities, as expected. Moreover, evaluated at zero both reach the expected values, $e^{-\lambda(0)}=1$ and $e^{\nu(0)}=const$.  Even more, the metric functions coincide at $r\approx 0.1$ so that, from the matching conditions with the Schwarzschild exterior, it corresponds to the radius of the surface of the star. In this regard, $e^{\nu(0.1)}=e^{-\lambda(0.1)}\approx0.46=1-2M/R$, form where the compactness of the star is
\begin{eqnarray}
\frac{M}{R}\approx 0.27.
\end{eqnarray}

\begin{figure}[h!]
\centering
\includegraphics[scale=0.5]{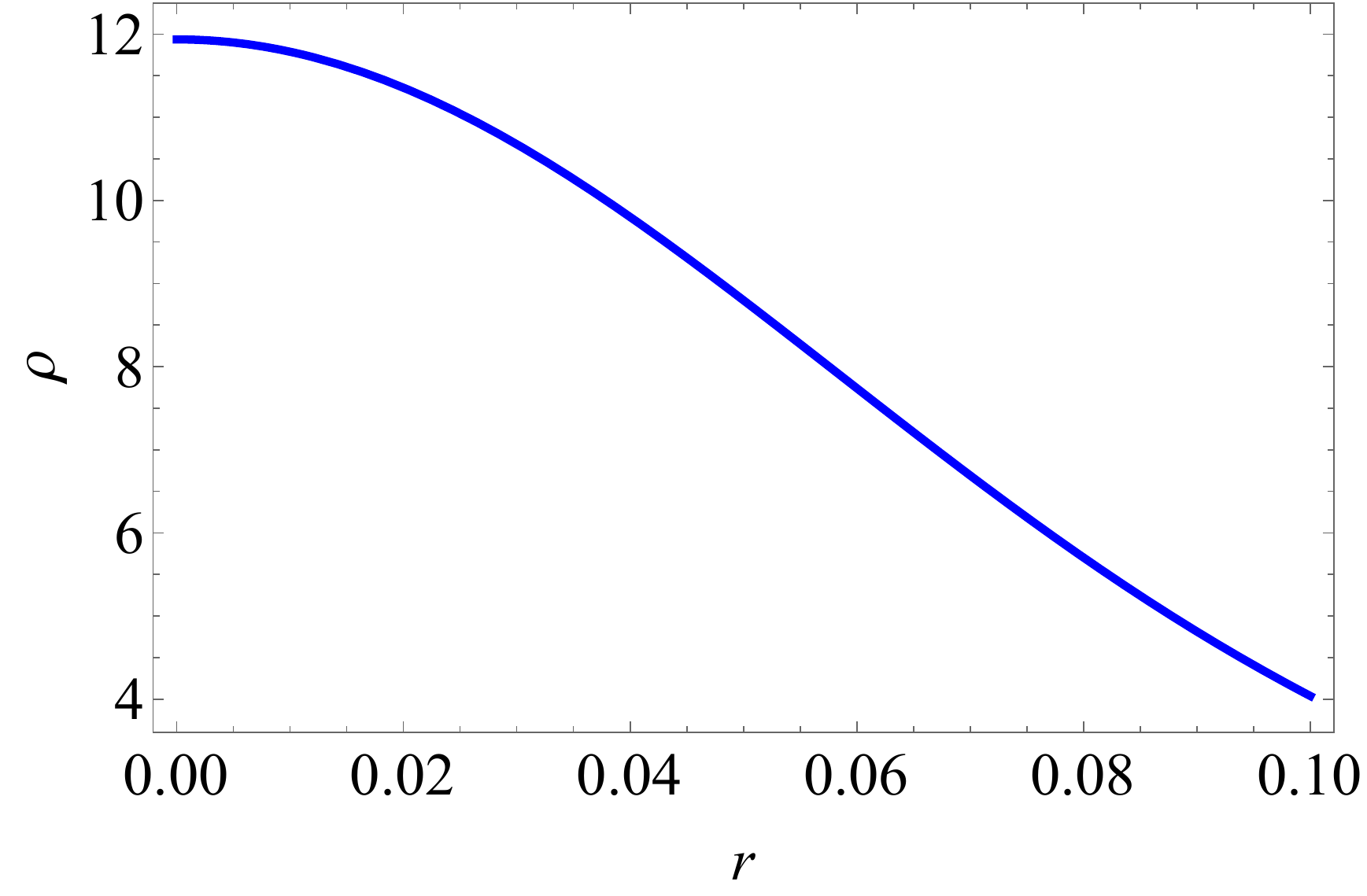}
\caption{\label{densidad-prcero}
Density $\rho$ as a function of the radial coordinate $r$ for the vanishing-complexity and $P_r =0$ model.
}
\end{figure}

\begin{figure}[h!]
\centering
\includegraphics[scale=0.5]{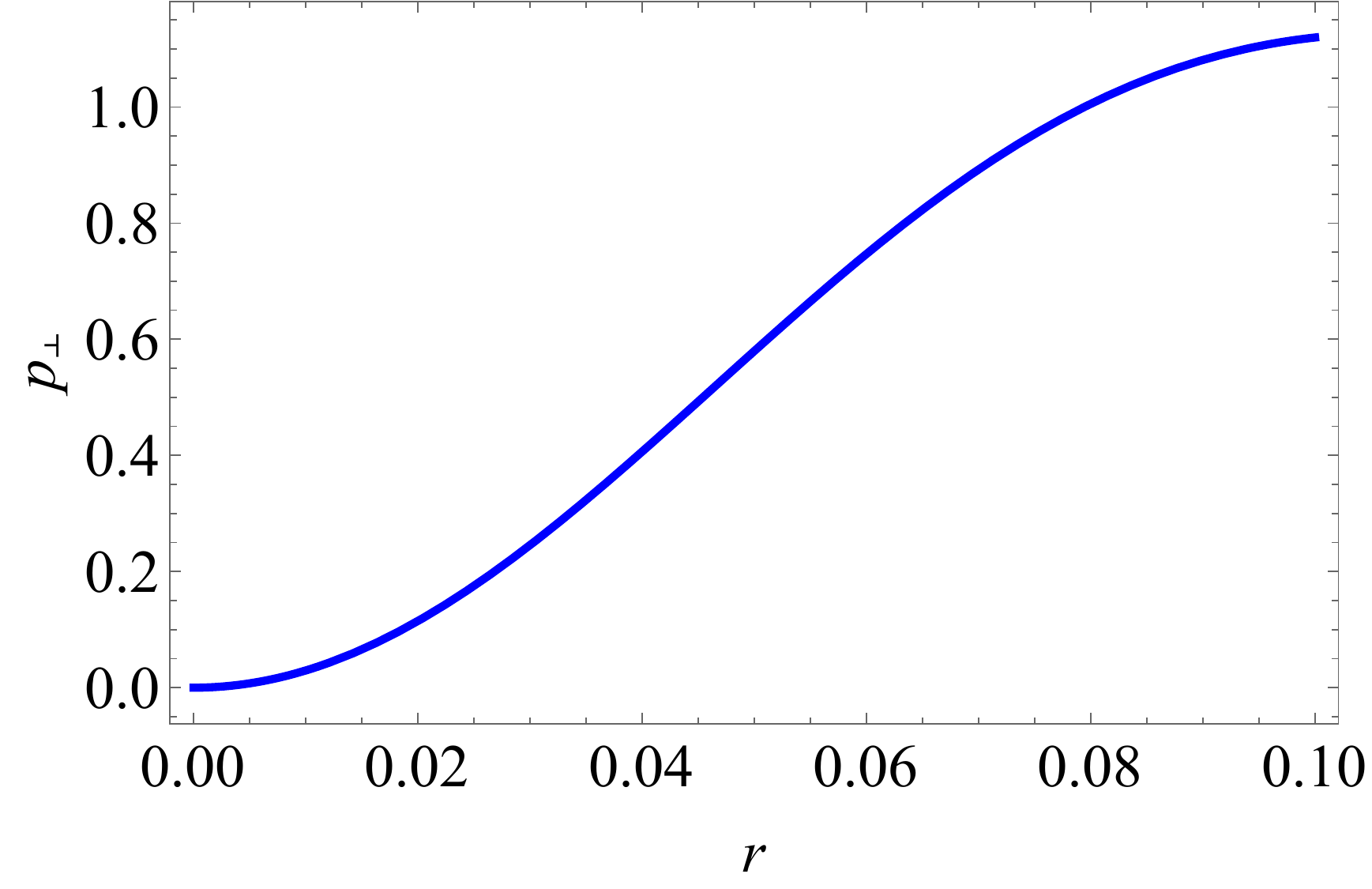}
\caption{\label{ptprcero}
Tangential pressure $P_{\perp}$ as a function of the radial coordinate $r$ for the vanishing-complexity and $P_r =0$ model.
}
\end{figure}

In figures \ref{densidad-prcero}
and \ref{ptprcero} we show the behavior of the matter sector. On the one hand, note that the density is positive inside the star, reaches its maximum at the centre and decreases monotonously  outwards, as expected. On the other hand we observe a behavior related to the well known Florides solution \cite{florides1974new}, where a system with no radial stresses has a positive tangential pressure that increases outwards just to keep the sphere stability. Additionally, note that the DEC is satisfied since $\rho$ $\geq$ $P_{\perp}$ and obviously $\rho$ $\geq$ $P_{r}$.

\begin{figure}[h!]
\centering
\includegraphics[scale=0.5]{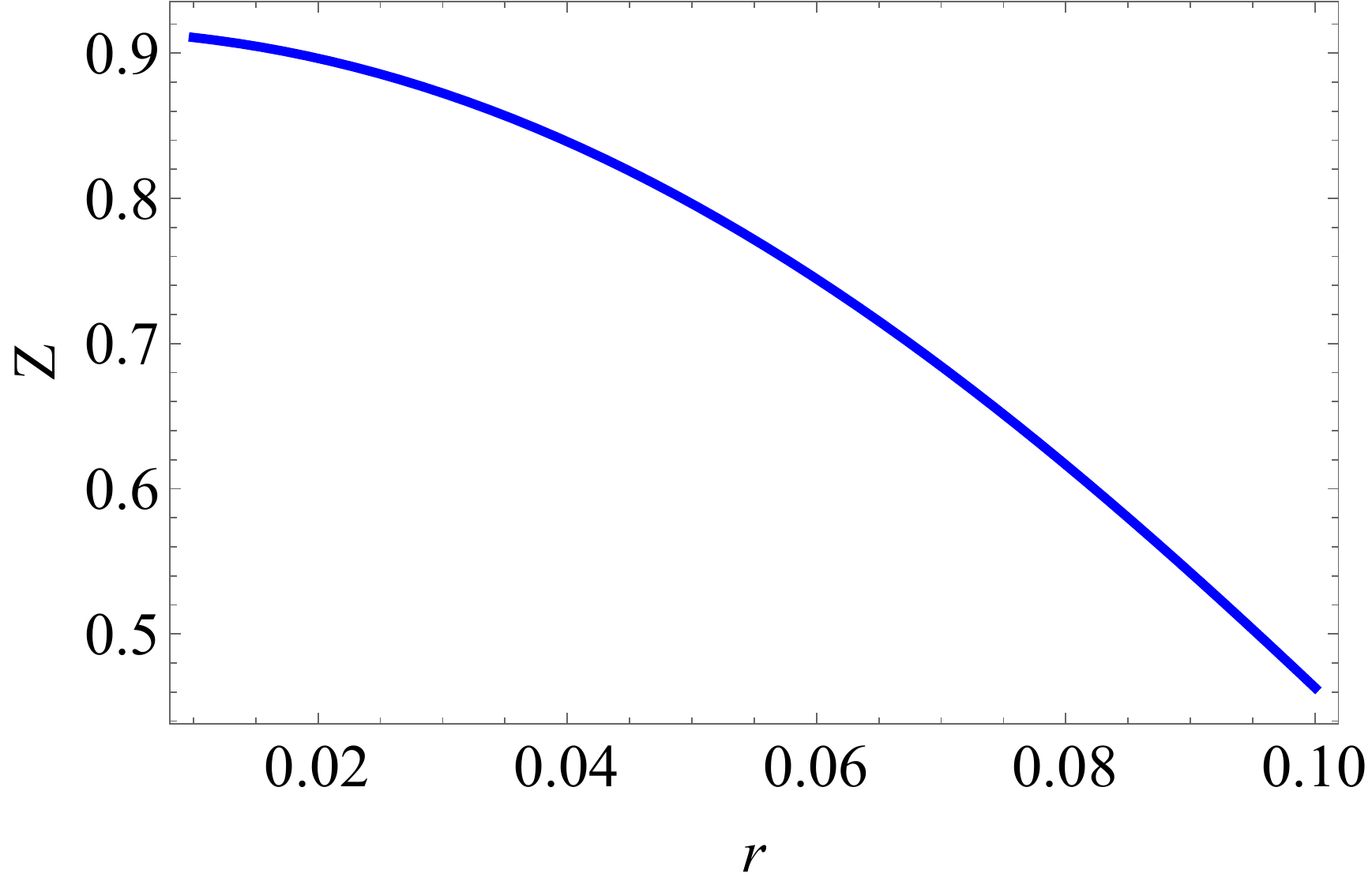}
\caption{\label{zprcero}
Redshift as a function of the radial coordinate $r$ for the vanishing-complexity and $P_r =0$ model.
}
\end{figure}

\newpage
In figure \ref{zprcero} it is shown the redshift as a function of the radial coordinate $r$ for the vanishing-complexity $P_r =0$ model and its observed that decreases as long as the radius increases. Even more, at the surface we have $Z(0.1)\approx 0.46$ which is less that the upper bound for the redshift at the surface of an anisotropic model satisfying the DEC condition corresponding to $Z_{bound}=5.11$.

\begin{figure}[h!]
\centering
\includegraphics[scale=0.5]{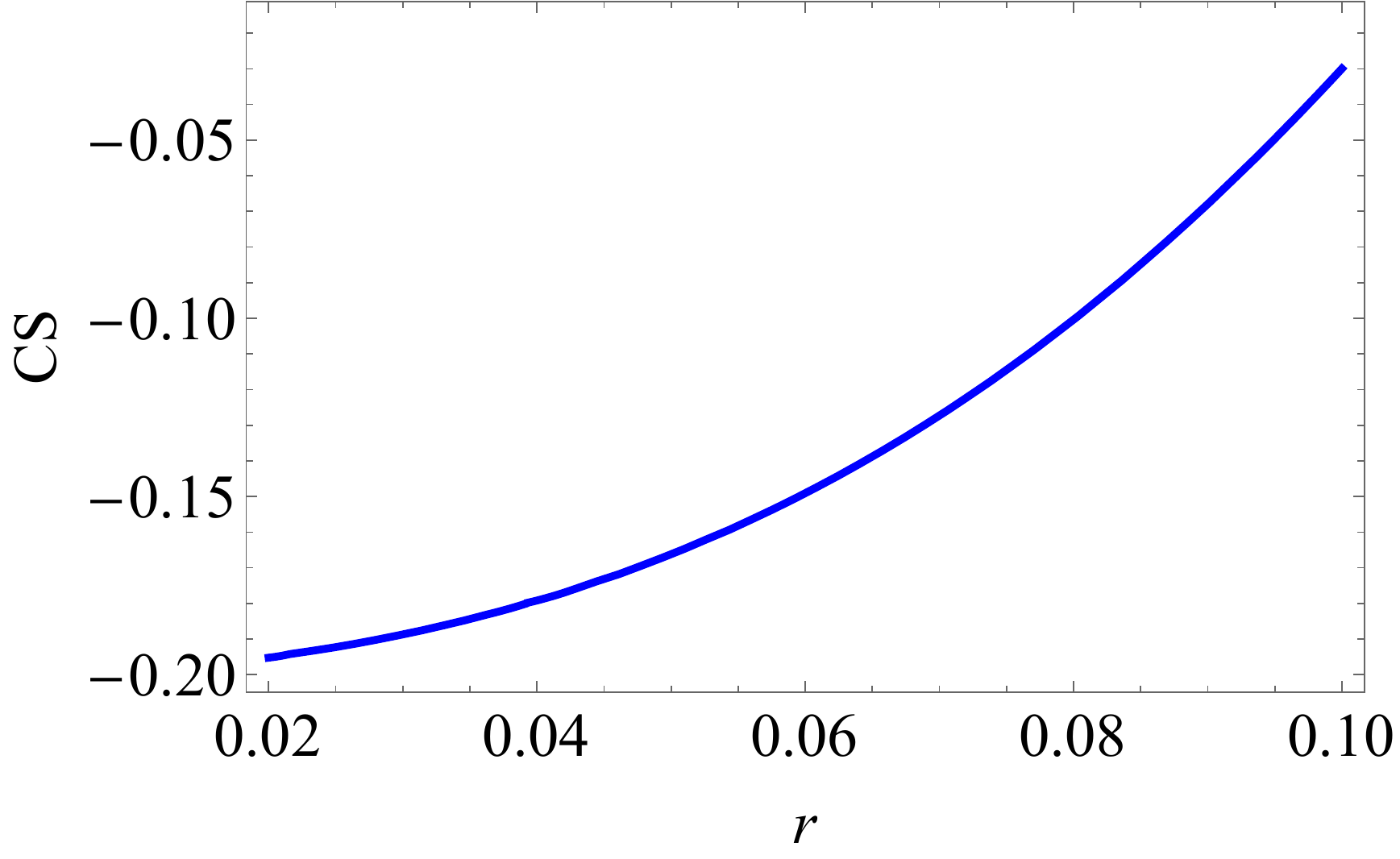}
\caption{\label{crack-st-prcero}
Cracking stability ($CS$) $=$ $\frac{dP_{\perp}}{d\rho}-\frac{dP_{r}}{d\rho}$ as a function of the radial coordinate $r$ for the vanishing-complexity and $P_r =0$ model.
}
\end{figure}

In figure \ref{crack-st-prcero}
we show the cracking stability as a function of $r$ and we note that this solution turns out to be stable against cracking since the CS curve lies in the range (-1, 0).

\subsection{Model II: vanishing-complexity polytrope}
The study of polytropes for anisotropic matter has been considered in detail in recent papers and play an important role in astrophysics \cite{herrera2013general, 3abellan2020general, herrera2014conformally, Ramos:2021drk}. In this model,
after adopting the polytropic equation of state in the case of anisotropic matter, we assume the vanishing of the complexity factor as a complementary condition. This model was proposed and briefly discussed in  \cite{herrera2018new}, however, we present here a more detailed analysis of the solution for this system. 

Let us start with the following conditions:
\begin{equation}
    P_r = K \rho^{\gamma} = K \rho^{1+\frac{1}{n}}\; ; \quad Y_{TF}=0\; ,
    \label{politropo}
\end{equation}
where constants $K$, $\gamma$, and $n$ are usually called the polytropic constant, polytropic exponent, and polytropic index, respectively. 
Now, the polytropic equations of state lead to \cite{herrera2018new} 
\begin{eqnarray}
&&\xi^{2}\frac{d\psi}{d\xi}\left[\frac{1-2(n+1)\alpha v/\xi}{1+\alpha\psi}\right]+v+\alpha\xi^{3}\psi^{n+1}\nonumber\\
&& 
+\frac{2\Pi\psi^{-n}\xi}{P_{rc}(n+1)}
\left[
\frac{1-2(n+1)\alpha v/\xi}{1+\alpha\psi}
\right]=0, \label{lem1}\\
&&\nonumber\\
&&\frac{dv}{d\xi}=\xi^{2}\psi^{n},
\label{lem2}
\end{eqnarray}
where
\begin{eqnarray}
\alpha = \frac{P_{rc}}{\rho_{c}}\; , \quad r = \frac{\xi}{A} \; , \quad A^2 = \frac{4 \pi \rho_{c}}{\alpha (n+1)} 
\end{eqnarray}
\begin{eqnarray}
\Psi^n = \frac{\rho}{\rho_{c}}\; , \quad v(\xi) = \frac{m(r)A^3}{4\pi\rho_{c}},
\end{eqnarray}
and subscript $c$ indicates that the quantity is evaluated at the center. At the boundary surface $r = r_\Sigma$ ($\xi = \xi_{\Sigma}$) we have $\Psi (\xi_{\Sigma}) = 0$.  These equations (\ref{lem1}) and (\ref{lem2}) form a system of two first order differential equations for the three unknown functions $\Psi$, $v$ and $\Pi$, so the third equation arises from the vanishing complexity condition ($Y_{TF}=0$) which can be writing as
\begin{eqnarray}
\frac{6\Pi}{n\rho_{c}}+\frac{2\xi}{n\rho_{c}}\frac{d\Pi}{d\xi}=\psi^{n-1}\xi\frac{d\psi}{d\xi}.
\end{eqnarray}

\begin{figure}[ht!]
\centering
\includegraphics[scale=0.5]{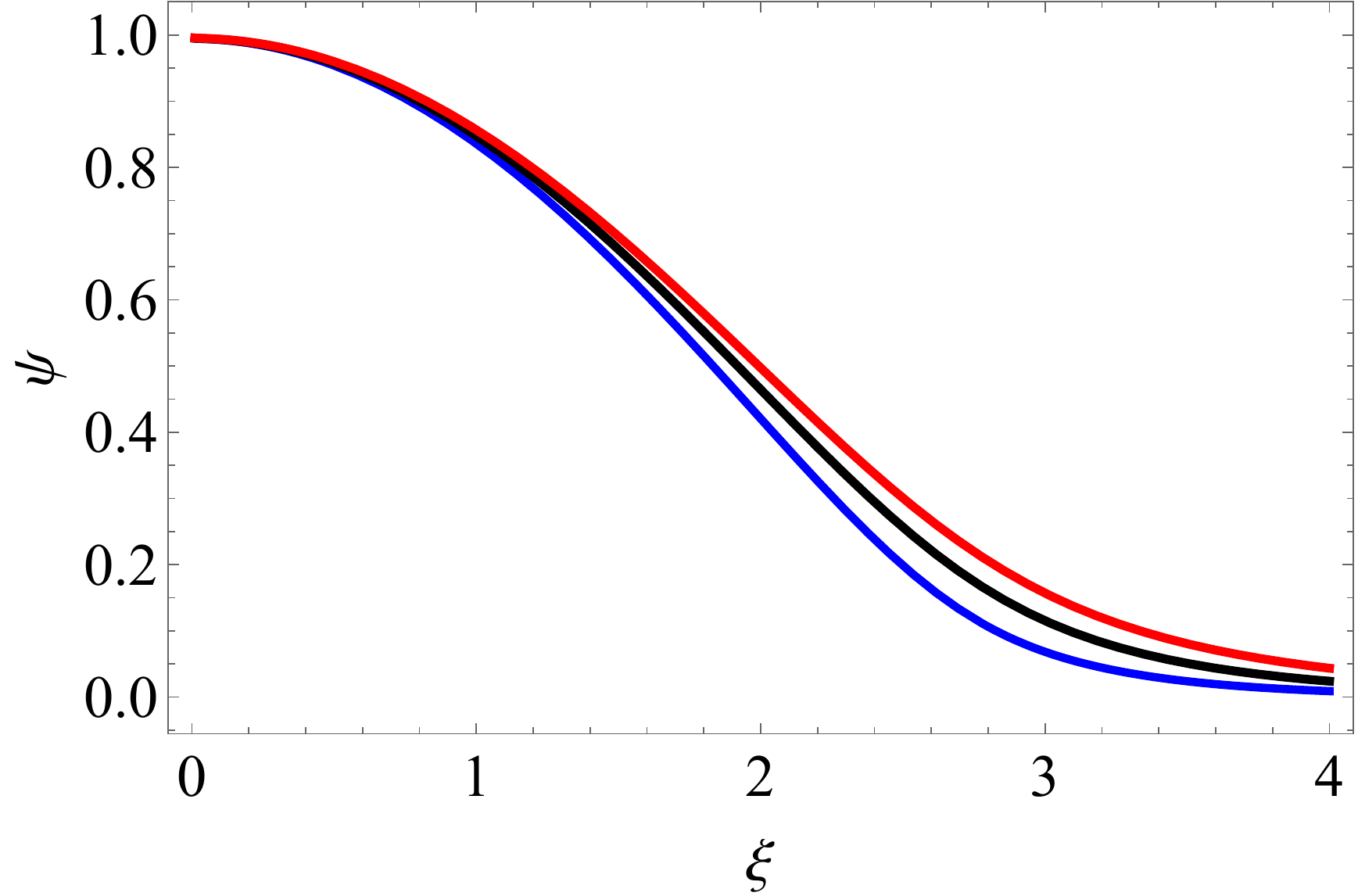}
\caption{\label{psipoli}
 $\psi$ as a function of $\xi$ for
 $\alpha=0.1$, $\rho_{c}=1$ and $n=0.3$ (blue line), $n=0.4$ (black line) and $n=0.5$ (red line) for the vanishing complexity polytrope.
}
\end{figure}

\begin{figure}[h!]
\centering
\includegraphics[scale=0.5]{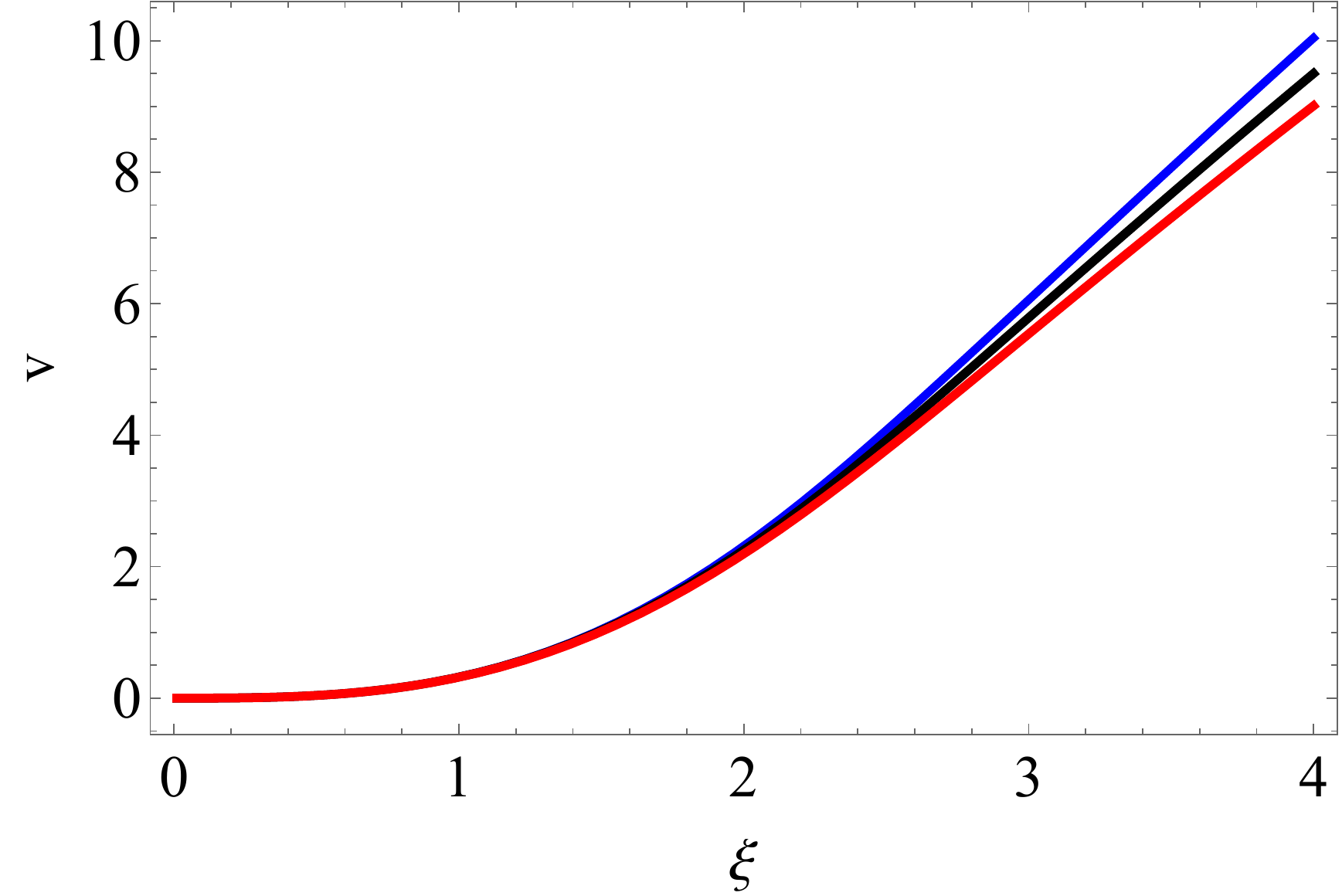}
\caption{\label{vpoli}
 $v$ as a function of $\xi$ for
 $\alpha=0.1$, $\rho_{c}=1$ and $n=0.3$ (blue line), $n=0.4$ (black line) and $n=0.5$ (red line) for the vanishing complexity polytrope.
}
\end{figure}

\begin{figure}[h!]
\centering
\includegraphics[scale=0.5]{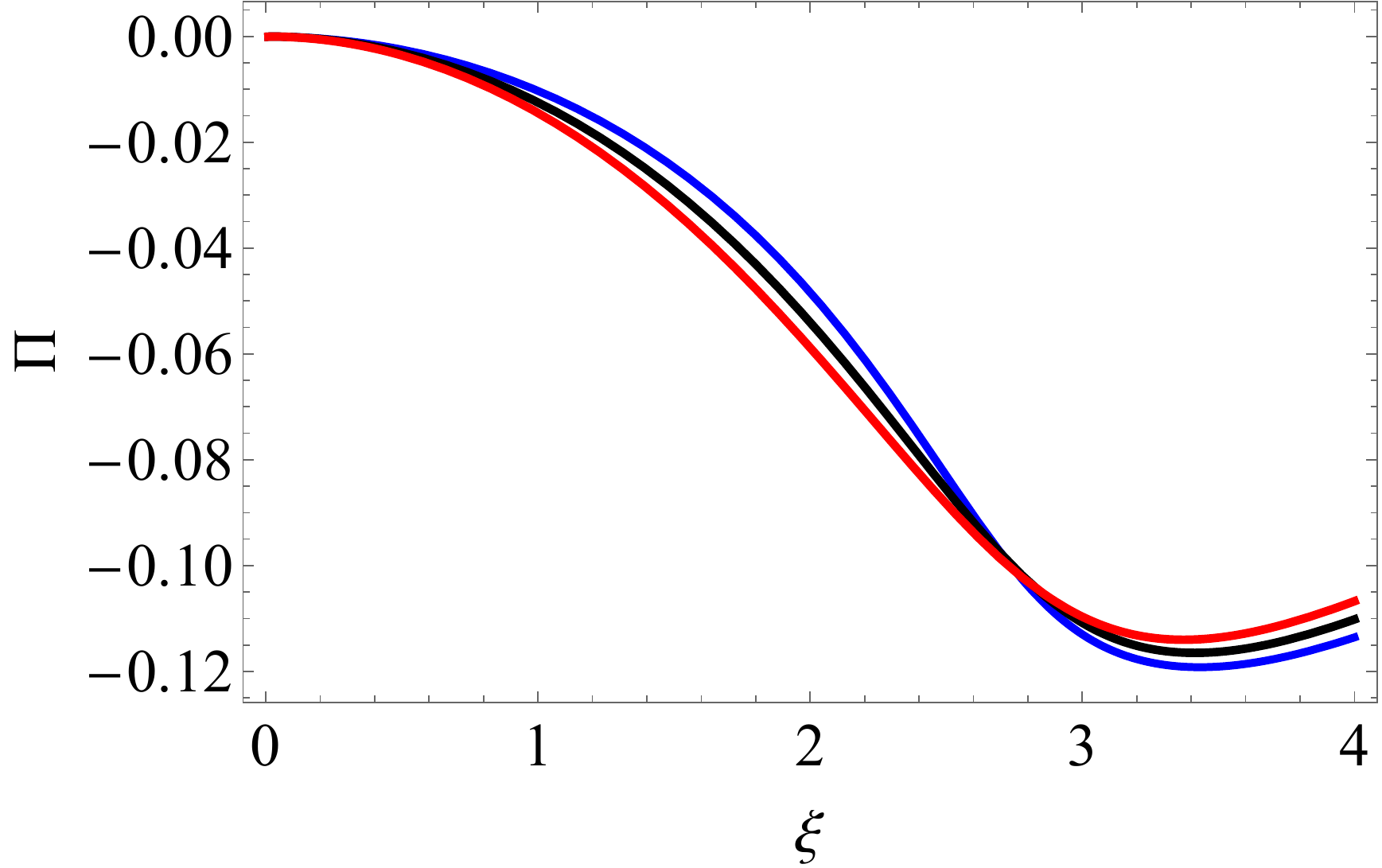}
\caption{\label{pipoli}
 $\Pi$ as a function of $\xi$ for
 $\alpha=0.1$, $\rho_{c}=1$ and $n=0.3$ (blue line), $n=0.4$ (black line) and $n=0.5$ (red line) for the vanishing complexity polytrope.
}
\end{figure}

In figures \ref{psipoli}, \ref{vpoli}  and
\ref{pipoli} we show the behavior of $\{\psi,v,\Pi\}$ as a function of the redefined variable ($\xi$) for different values of the duplet of parameters ($n$, $\alpha$) indicated in the caption of the corresponding figures. We observe in figure \ref{psipoli} that the function $\Psi$ (energy density) is monotonously decreasing, as expected for well behaved general relativistic polytropes. Furthermore the numerical integration has been stopped when $\psi$ has a root which, for the conditions imposed here, occurs around $\xi\approx4$.
Moreover, in figure \ref{vpoli} we can observe a monotonously increasing correct behavior for the mass function, this already allows us to expect an adequate behavior of the metric function $\lambda$. 

\begin{figure}[ht!]
\centering
\includegraphics[scale=0.5]{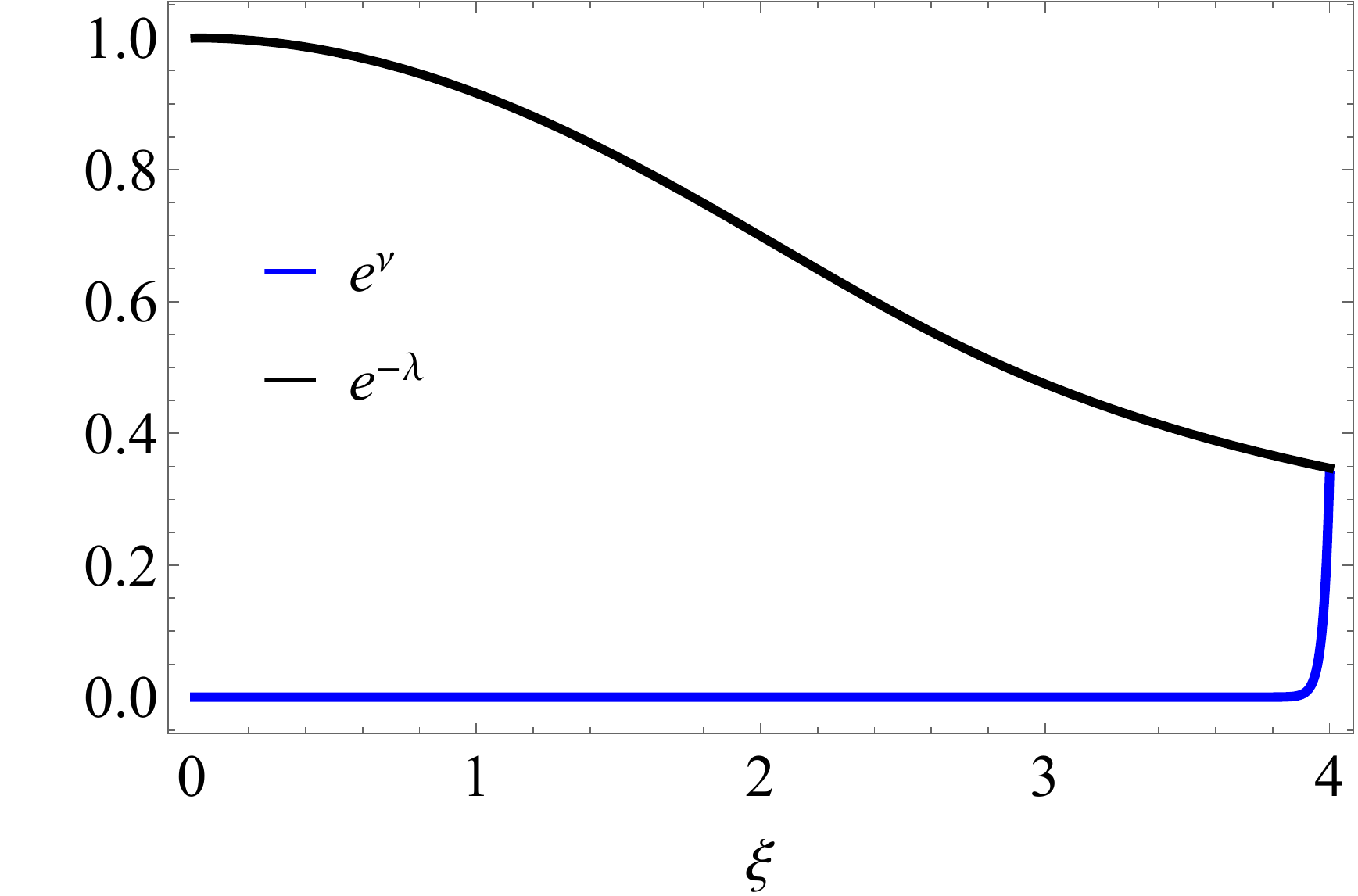}
\caption{\label{metricaspoli}
 $e^{\nu}$ (blue line), $e^{-\lambda}$ (black line) as a function of $\xi$ for
 $\alpha=0.1$, $\rho_{c}=1$, $K=0.9$ and $n=0.3$ for the vanishing complexity polytrope.
}
\end{figure}

Figure \ref{pipoli} is dedicated to exposing the behavior of the local anisotropy of the pressure as a function of the dimensionless parameter $\xi$. For this anisotropy function we obtain the usual  behavior.
In figure \ref{metricaspoli} it is shown the geometric sector. Observe that the metric functions are positive, finite and free of singularities, as it should be. Besides, evaluated at zero both reach the expected values, $e^{-\lambda(0)}=1$ and $e^{\nu(0)}=const$ and both functions coincide at $\xi\approx 4$ which define the surface of the star and allows us to compute the compactness $M/R$. For example, for $n=3$ we obtain $e^{\nu(4)}=e^{-\lambda(4)}\approx0.3473=1-2M/R$ form where, the compactness reads
\begin{eqnarray}
\frac{M}{R}\approx0.326318 .
\end{eqnarray}

\begin{figure}[ht!]
\centering
\includegraphics[scale=0.5]{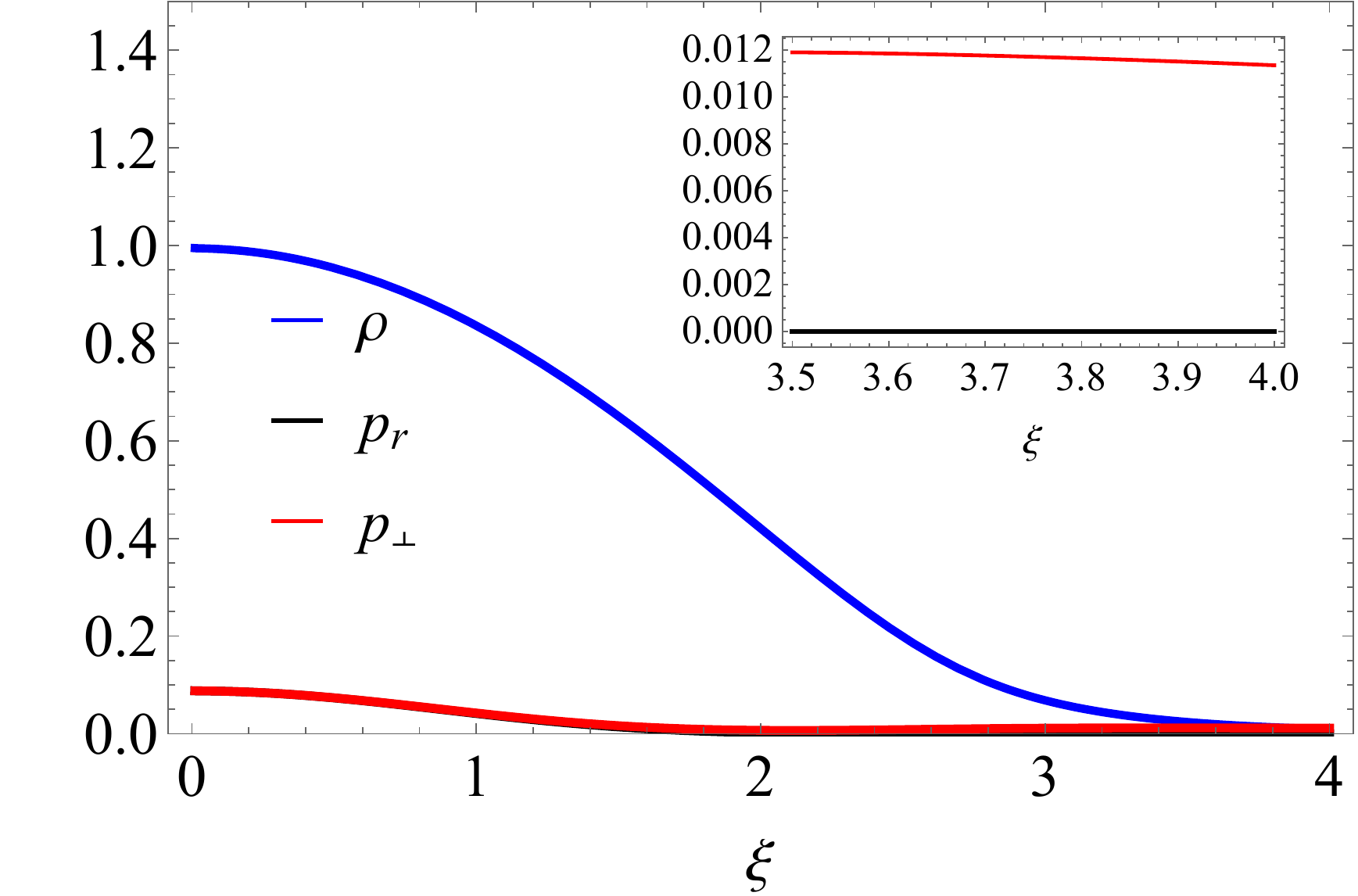}
\caption{\label{matterpoli}
 $\rho$ (blue line), $p_{r}$ (black line) and $p_{\perp}$ (red line) as a function of $\xi$ for
 $\alpha=0.1$, $\rho_{c}=1$, $K=0.9$ and $n=0.3$ for the vanishing complexity polytrope. 
}
\end{figure}

\begin{figure}[ht!]
\centering
\includegraphics[scale=0.5]{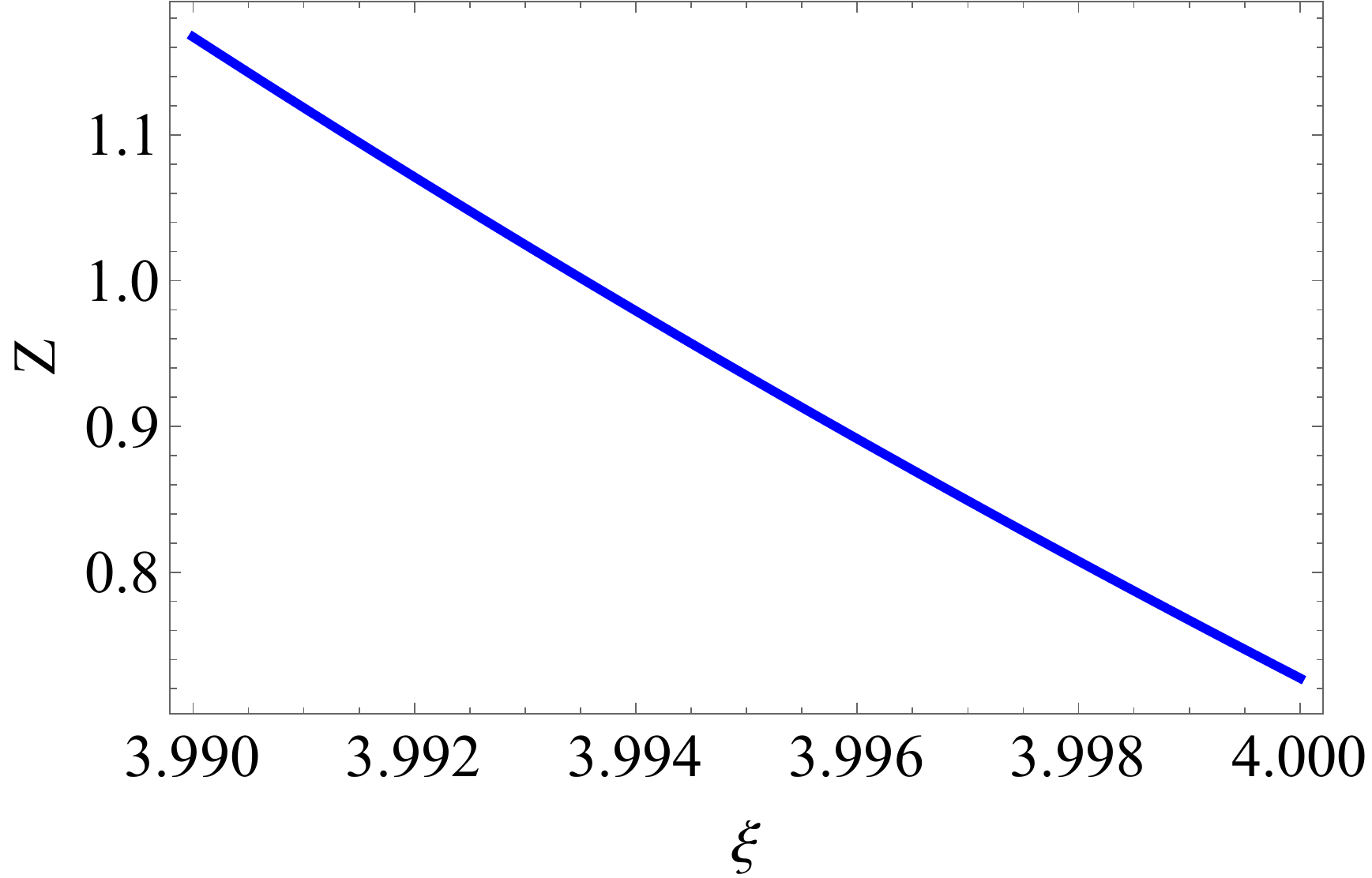}
\caption{\label{zpoli}
 Redshift  as a function of $\xi$ for $\alpha=0.1$, $\rho_{c}=1$, $K=0.9$ and $n=0.3$ for the vanishing complexity polytrope.
}
\end{figure}

\begin{figure}[ht!]
\centering
\includegraphics[scale=0.5]{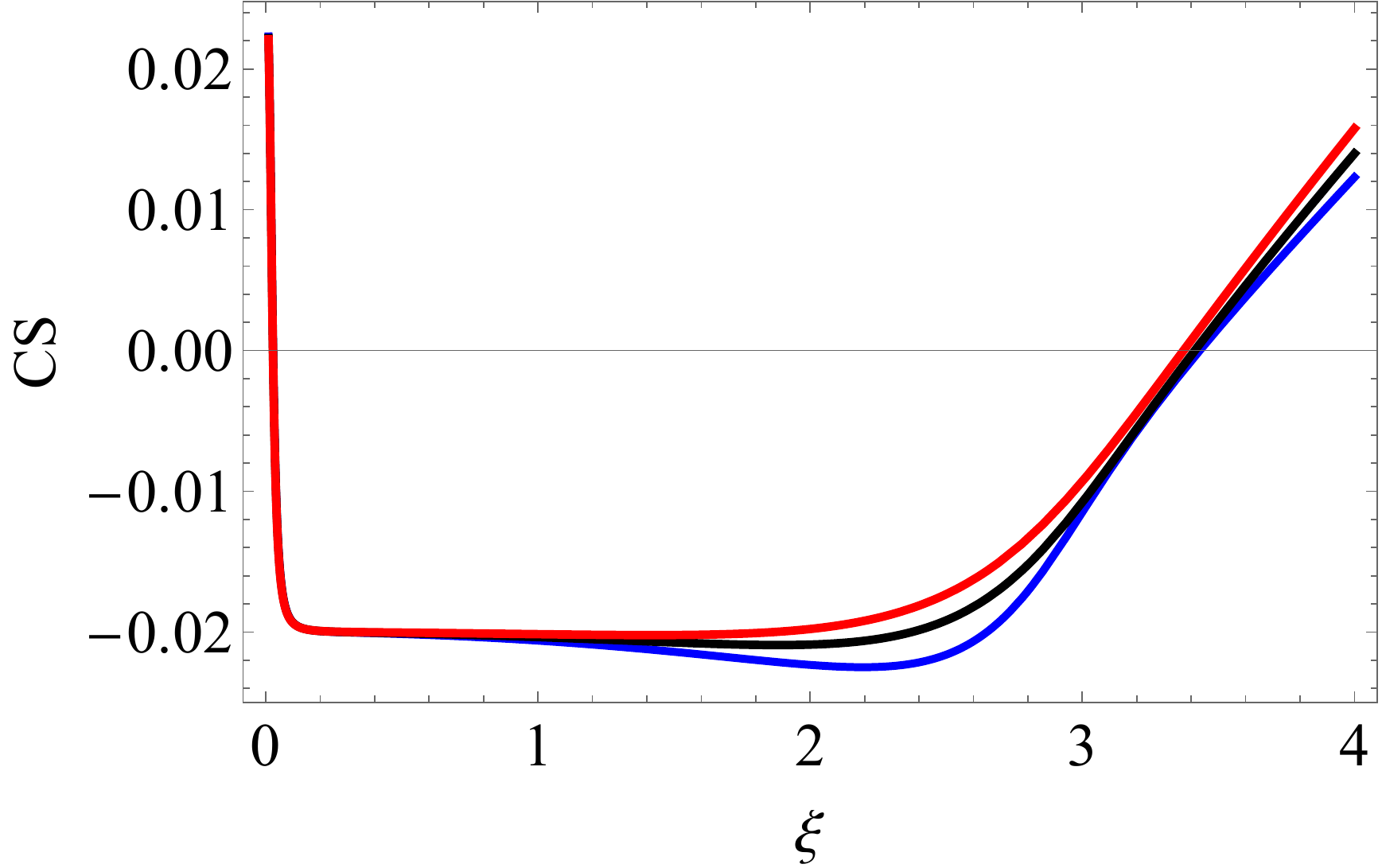}
\caption{\label{crack-st-poli}
 Cracking stability ($CS$) $=$ $\frac{dP_{\perp}}{d\rho}-\frac{dP_{r}}{d\rho}$ for $K=0.9$, $\alpha=0.1$, $\rho_{c}=1$ and  $n=0.3$ (blue line), $n=0.4$ (black line) and $n=0.5$ (red line) for the vanishing complexity polytrope.
}
\end{figure}

In figure \ref{matterpoli} we show the matter sector (hydrodynamic variables) plotted as a function of the dimensionless radial coordinate $\xi$ for different choices of the parameters shown in the legend of the figure. Note that the density, radial pressure and tangential pressure are positive quantities inside the star, they reach their maximum at the centre and then monotonously decrease outwards, representing an appropriate behavior. 
Besides, for this model the DEC is satisfied as it can be appreciated in the same figure, where the density is greater than both pressures, radial and tangential. In addition, from the same figure the fact that the local anisotropy of pressure has the appropriate behavior is immediately appreciated.
In figure \ref{zpoli} the redshift is represented as function of the normalized quantity $\xi$. For this model, our fifth physical condition is satisfied since the redshift is a decreasing function when the dimensionless radius increases and its value at the surface is less than the universal bound for anisotropic solutions satisfying the DEC. For example, for $n=3$ the redshift at the surface is given by $Z(4)\approx 0.727326<Z_{bound}$, as expected. 
In figure \ref{crack-st-poli} we show the cracking stability for the parameters involved in the legend.
The plot shows that all the models under consideration are unstable against cracking because the condition takes positive values for small radius, which is not allowed. There are several works that carry out studies of cracking presented by different types of relativistic polytropes, so this represents an expected result as it was referred in \cite{herrera2016cracking,contreras2021gravitational}, to give some examples. In \cite{contreras2021gravitational} the possible relationship between cracking and complexity was even raised.

\subsection{Model III: Vanishing complexity with a non--local equation of state}
The system with a non--local equation of state was first proposed in \cite{Hernandez_2004} and is given by the expression
\begin{eqnarray}
P_{r}=\rho-\frac{2}{r^{3}}\int\limits_{0}^{r}\tilde{r}^{2}\rho(\tilde{r})d\tilde{r}+\frac{c}{2\pi r^{3}},
\end{eqnarray}
or, by using the definition of the mass function, we get
\begin{eqnarray}
P_{r}=\frac{m'}{4\pi r^{2}}-\frac{m}{2\pi r^{3}}+\frac{c}{2\pi r^{3}}.
\label{noloc}
\end{eqnarray}
Now, using the expression (\ref{noloc}) in conjunction with the condition for vanishing of the complexity factor (\ref{vanishingYTF}) we have the necessary tools to solve Einstein's equations. For this last model, we found physically acceptable solutions only for $c=0$. 

\begin{figure}[ht!]
\centering
\includegraphics[scale=0.5]{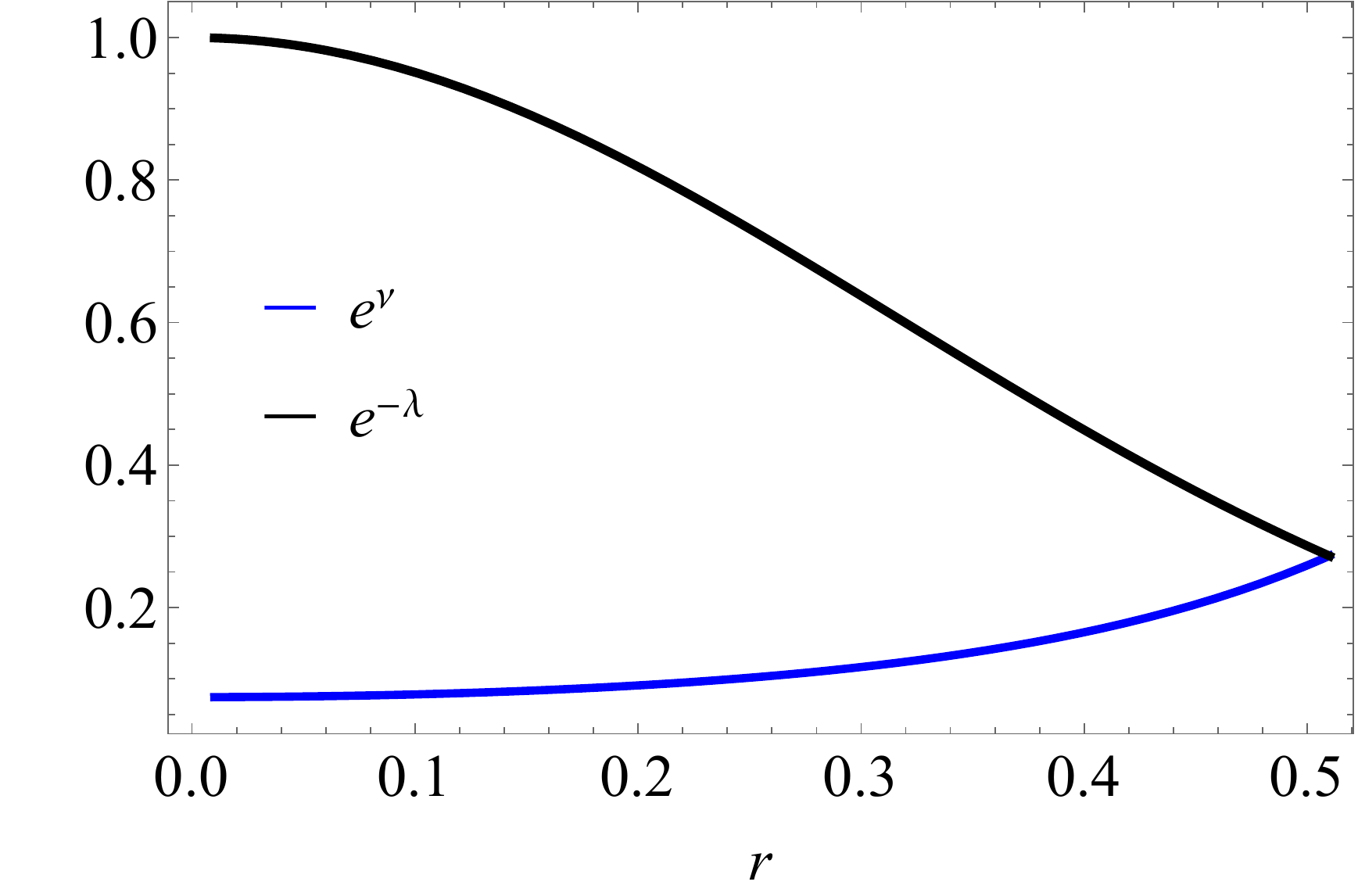}
\caption{\label{matricas-nolocal}
 $e^{\nu}$ (blue line), $e^{-\lambda}$ (black line) function of $r$ for the vanishing complexity non--local equation of state model with $c=0$.
}
\end{figure}

In figure \ref{matricas-nolocal} we show the 
behavior of the metric functions, with respect to the radial coordinate $r$, which are positive, finite and free of singularities, as they should be for a physical accepted solution. Furthermore, evaluated at zero, both reach the expected values, $e^{-\lambda(0)}=1$ and $e^{\nu(0)}=const$. For this model we find $e^{\nu(0.5)}=e^{-\lambda(0.5)}=1-2M/R$, from where
\begin{eqnarray}
\frac{M}{R}\approx 0.370381 .
\end{eqnarray}

\begin{figure}[ht!]
\centering
\includegraphics[scale=0.5]{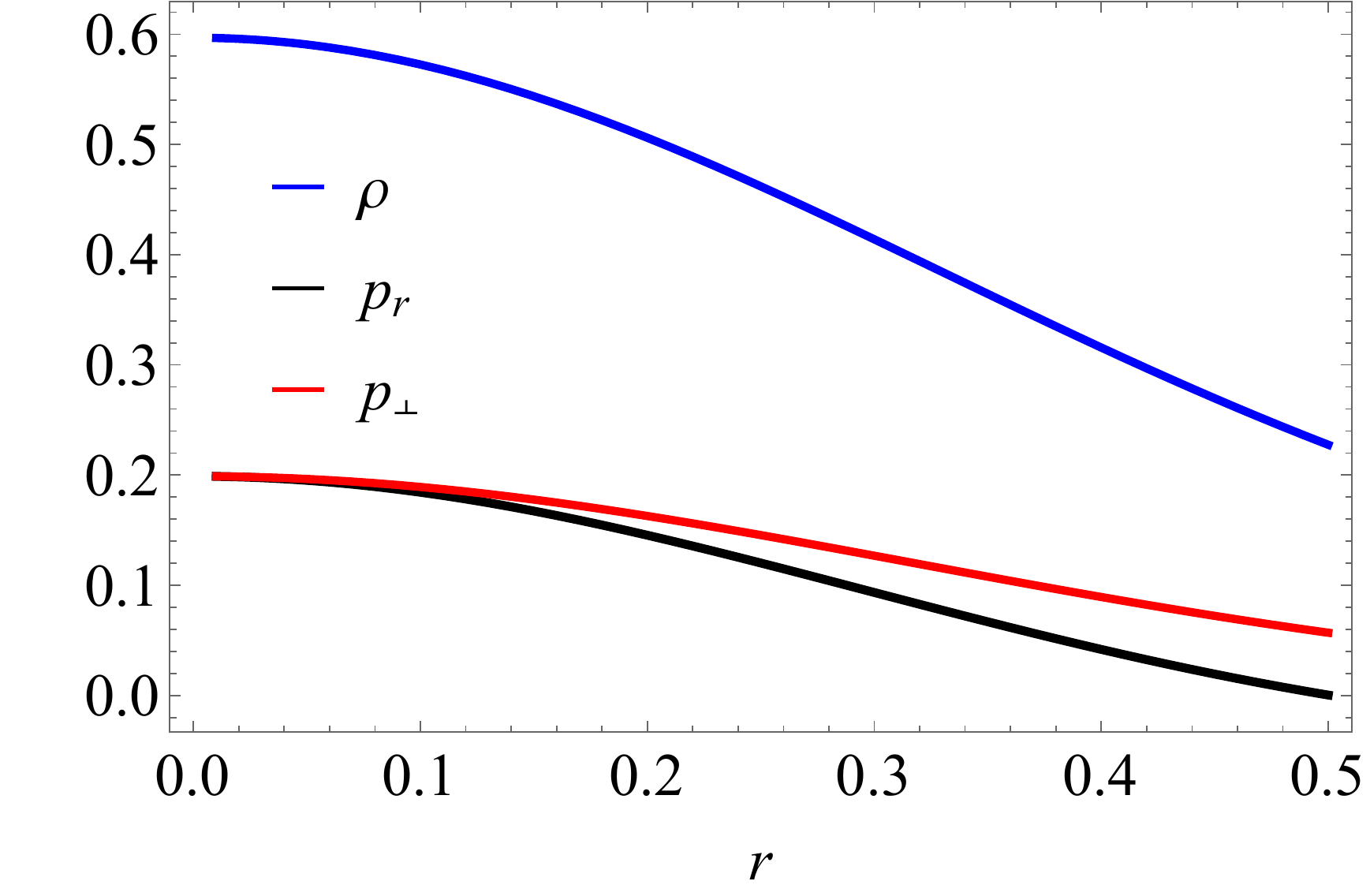}
\caption{\label{matter-nolocal}
 $\rho$ (blue line), $P_{r}$ (black line) 
 and $P_{\perp}$ (red line) as a function of $r$ for the vanishing complexity non--local equation of state model with $c=0$.
}
\end{figure}

In figure \ref{matter-nolocal} the matter sector and the fluid tensions are shown. Note that the density and pressure (radial and tangential) behave as expected: positive quantities inside the star, their maximum is at the centre and then decrease monotonously outwards. For this last model, the DEC is also satisfied since $\rho$ $\geq$ $P_{r}$, and $\rho$ $\geq$ $P_{\perp}$ and also the anisotropy will have the desired behavior.   

\begin{figure}[ht!]
\centering
\includegraphics[scale=0.5]{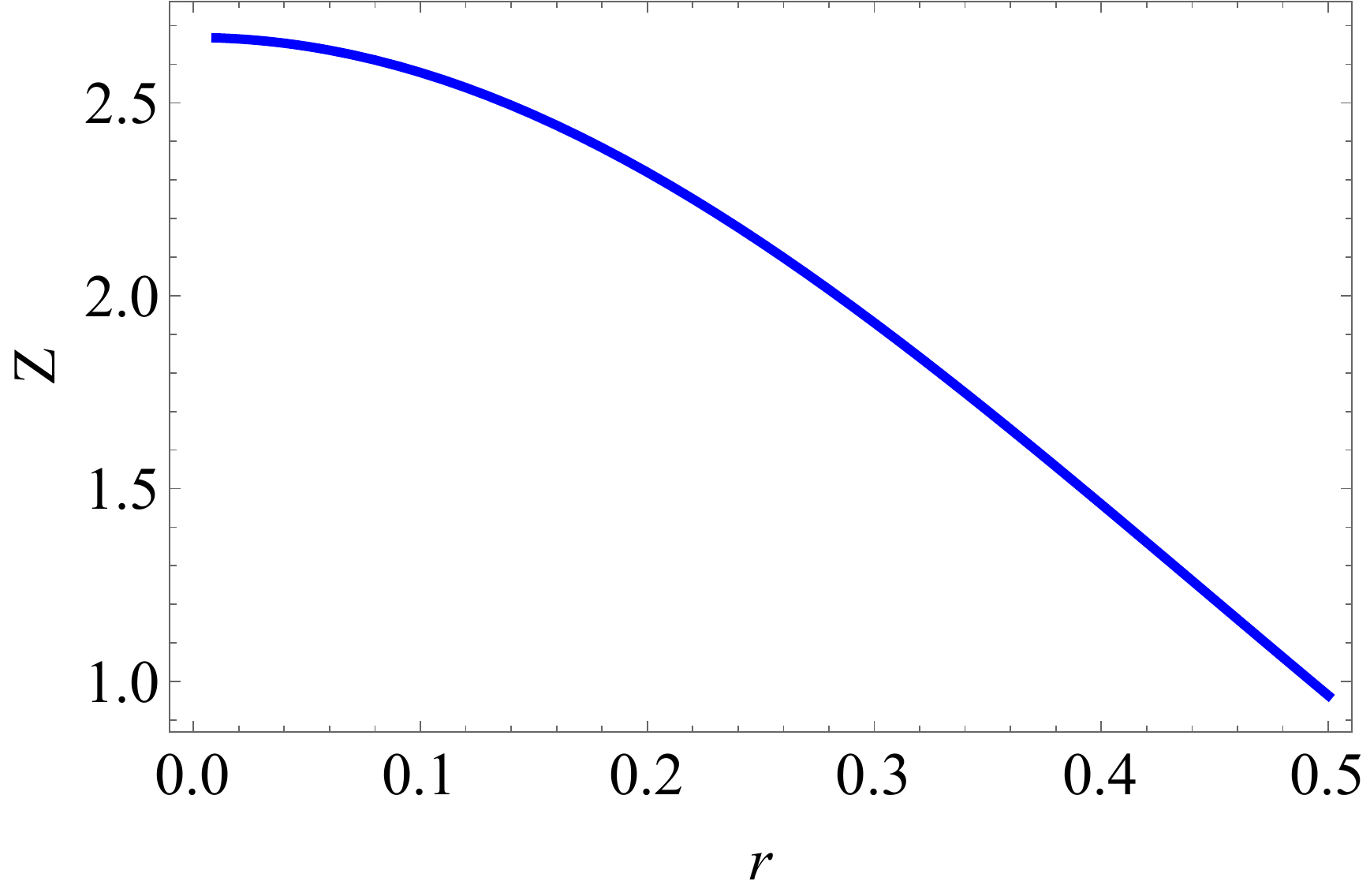}
\caption{\label{znolocal}
 Redshift as a function of $r$ for the vanishing complexity non--local equation of state model with $c=0$.
}
\end{figure}

\begin{figure}[ht!]
\centering
\includegraphics[scale=0.5]{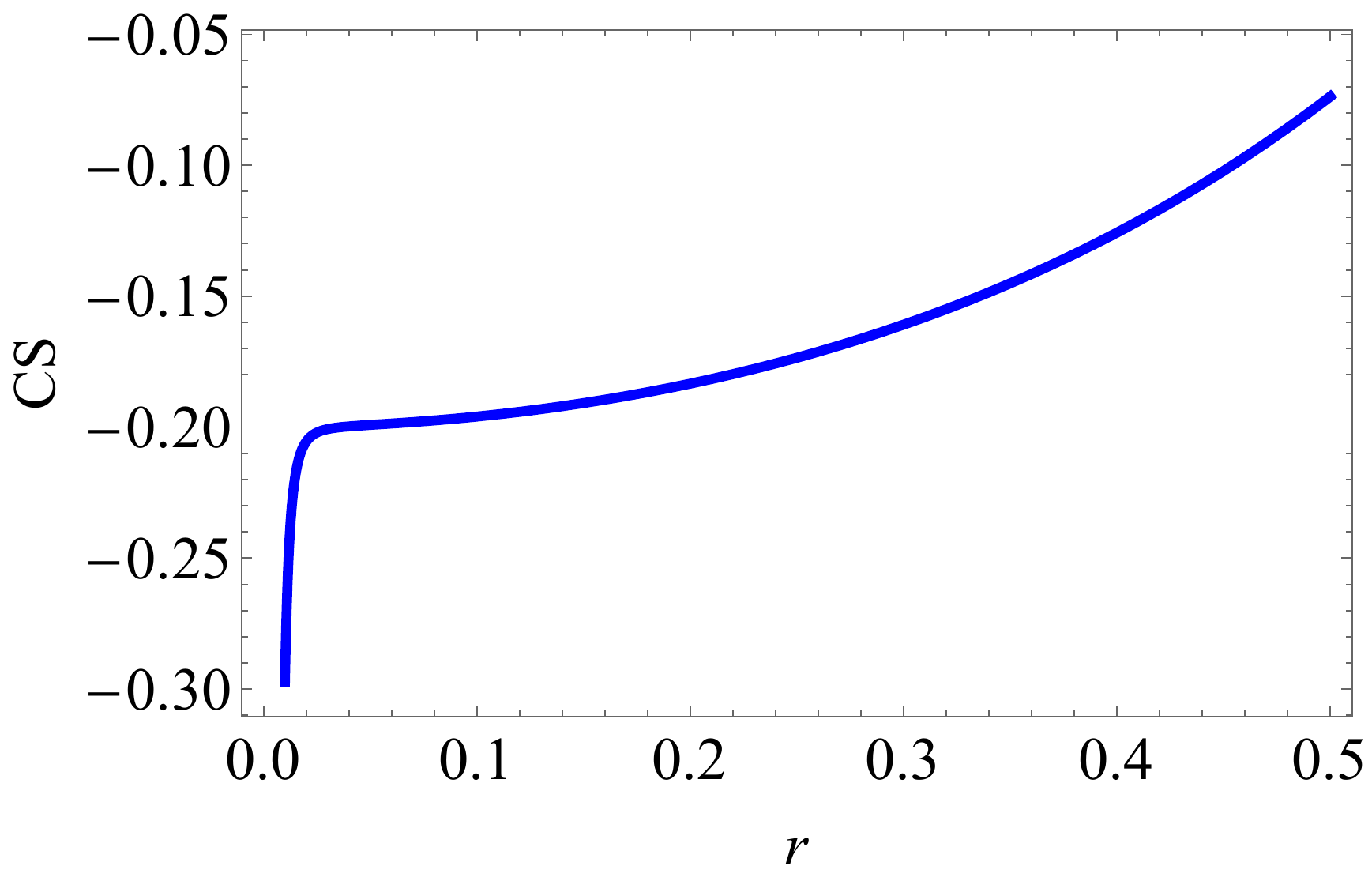}
\caption{\label{crack-st-nonlocal}
 $CS=$ $\frac{dP_{\perp}}{d\rho}-\frac{dP_{r}}{d\rho}$  as a function of $r$ for the vanishing complexity non--local equation of state model with $c=0$.
}
\end{figure}

In figure \ref{znolocal}
we show the behavior of the redshift as a function of $r$ for the non--local equation of state with $c=0$. Note that the function decreases as long as the radius increases and $Z(0.5)\approx 0.964038<Z_{bound}$, as expected. Finally, in figure \ref{crack-st-nonlocal}, we show the stability condition against cracking. Note that the function $CS=$ $\frac{dP_{\perp}}{d\rho}-\frac{dP_{r}}{d\rho}$  lies in the range $(-1,0)$ so the solution is stable against cracking (or overturning). 

\section{Final Remarks}

A new concept of complexity, for static spherically symmetric relativistic fluid distributions, was recently defined in \cite{herrera2018new} which arises from the basic assumption that the less complex systems corresponds to an homogeneous (in the energy density) fluid distribution with isotropic pressure. Then, from the orthogonal splitting of the Riemann tensor, appears the structure scalar $Y_{TF}$ as an obvious candidate to measure the degree of complexity so we call it the complexity factor. This scalar function contains contributions from the energy density inhomogeneity and the local pressure anisotropy, combined in a very specific way so it measures the departure from the value of the Tolman mass (active gravitational mass) in the homogeneous and isotropic case. The complexity factor so defined, obviously vanishes for a homogeneous and isotropic fluid, where the two terms in (\ref{ytf}) vanish identically, but it also presents the wealth of vanishing itself when the two terms in (\ref{vanishingYTF}) cancel each other. This allows finding a great variety of models with zero complexity different from the trivial case. On that the general proposal of this work was based.\\

The Einstein field equations for a spherically symmetric static, anisotropic fluid form a system of three differential equations for five unknown quantities, the two metric functions $\nu $, $\lambda $ and the three hydrodynamic variables (matter sector) that describe the energy density $\rho$, radial $P_r$ and tangential pressure $P_{\perp}$ of the fluid. Accordingly, if we impose the condition $Y_{TF}=0$ we shall need still another condition in order to solve the system. In this work, together with the vanishing complexity condition (\ref{vanishingYTF}) we proposed the conditions $P_{r}=0$, a polytrope and a non-local equation of state so, we were able to build three models using the extra information needed for closing the system. We analyzed the plausibility of each one based on the physical conditions established for the existence of anisotropic compact star models. All the considered cases behaved as expected in terms of metric potentials, matter sector, energy conditions, redshift and stability against cracking, except for the vanishing complexity polytrope case where we found that it presents instability under cracking. This suggests that the vanishing of the complexity parameter for a self--gravitating polytropic compact object induces instability that could eventually fracture the star. This represents a well known result (see for example \cite{contreras2021gravitational}). On the other hand, for the non--local equation of state model we found physically acceptable solutions only if $c=0$, therefore suggesting that the value of $c$ is closely related to the complexity of the system.\\

It is worth mentioning that some extra models were also considered in the context of this work. One of them, was based on the equation of state $P_{\perp} = 0$ for all points inside the compact stellar fluid distribution. For this model, it was impossible to satisfy a realistic physical solution that matches with the vanishing complexity condition. This can be understood in a very simple way, because if we impose $P_{\perp} = 0$ for the tangential pressure in (\ref{vanishingYTF}), the resulting expression asserts that $P_{r}$ $\propto$ $\rho'$. So, if the energy density of the system corresponds to a well behaved decreasing function, its derivative will be a negative function, $\rho' < 0$. Thus, we are forced to have a model where the radial pressure is negative and this fact is clearly not satisfactory. It is worth mentioning that something similar seems to happen when trying to make the Cosenza anisotropic model \cite{cosenza1981some} consistent together with the vanishing complexity condition. Apparently the choice of the other extra condition (besides $Y_{TF}=0$), does not necessary turn to be compatible with the fact that the complexity of the system is zero. It would be interesting to study if some other value for the complexity factor (different from zero) make these systems coherent and physically acceptable when we use the conditions together. A last model, that is based on the implementation of both, vanishing complexity and confomally flat condition simultaneously was also considered. However, from
(\ref{ytf}) it is clear that conformally flat condition ($E=0$) reduces to vanishing complexity leading to the Schwarzschild interior.
\\

In summary, the models we have presented here satisfy most of the physicals conditions, so they are good candidates to represent realistic models with vanishing complexity. This represents a key that assert that we may be able to extend the notions of simplicity for self--gravitating objects and understand its implications in a more general and better way.


\bibliography{ref}
\bibliographystyle{unsrtnat}

\end{document}